\documentstyle[12pt] {article}
\def\zid{1\kern-0.36em\llap~1}

\newcommand{\beq}{\begin{equation}}
\newcommand{\ber}{\begin{eqnarray}}
\newcommand{\eeq}{\end{equation}}
\newcommand{\eer}{\end{eqnarray}}

\begin{document}

\begin{titlepage}
\rightline{SUNY BING 5/27/97R2, hep-ph/9706469}
\vspace{1mm}
\begin{center}
{\bf IMPORTANCE OF TESTS FOR THE COMPLETE LORENTZ
STRUCTURE \newline
OF THE $t \rightarrow W^+ b $ VERTEX AT HADRON COLLIDERS}\\
\vspace{2mm}
Charles A. Nelson\footnote{Electronic address: cnelson @
bingvmb.cc.binghamton.edu }, Brian T. Kress, Marco Lopes, and
Thomas P. McCauley\\
{\it Department of Physics, State University of New York at
Binghamton\\
Binghamton, N.Y. 13902-6016}\\[2mm]
\end{center}


\begin{abstract}

The most general Lorentz-invariant decay-density-matrix
for $t\rightarrow
W^{+}b\rightarrow (l^{+}\nu )b$, or for $t\rightarrow
W^{+}b\rightarrow (j_{\bar
d}j_u)b$, is expressed in terms of eight helicity parameters.
The parameters
are physically defined in terms of
partial-width-intensities for
polarized-final-states in \newline $t\rightarrow
W^{+}b$ decay.  The parameters
are the partial width, the $b$ quark's chirality parameter
$\xi$, the $W^+$ polarimetry parameter $\sigma$, a ``pre-SSB"
test parameter $\zeta$, and
four $W_{L}$-$W_{T}$ interference parameters
$\eta$, $\eta^{'}$, $\omega$, $\omega^{'}$ which test for
$\tilde T_{FS}$
violation.  They can be used to test for non-CKM-type CP
violation, anomalous
$\Gamma_{L,T}$'s, top weak
magnetism, weak electricity,
and second-class currents. By stage-two spin-correlation
techniques, percent level statistical \newline uncertainites
are
typical for
measurements at the Tevatron, and several mill level \newline
uncertainites are typical at the LHC.
\end{abstract}

\end{titlepage}

\section*{}

Historically in the study of the weak charged-current in
muonic and in hadronic processes, it has been important to
determine the ``complete Lorentz structure" directly from
experiment.  Here, we exploit the fact that the high mass,
$\sim 175$  $GeV$, of the newly discovered top quark [1,2,3]
implies that $t \rightarrow W^+ b$ decay allows for
probes of new physics beyond the standard model because this
decay is essentially free of complicating hadronization
effects, and that[4] approximately $70 \% $ of the final
$W$'s
will be longitudinally polarized, i.e. $\Gamma_L / \Gamma_T =
2.43$ if the standard model(SM) is indeed correct.

In the  $t$ rest frame, the
matrix element for $t \rightarrow W^{+} b$ is
\beq
\langle \theta _1^t ,\phi _1^t ,\lambda _{W^{+} } ,\lambda
_b
|\frac
12,\lambda _1\rangle =D_{\lambda _1,\mu }^{(1/2)*}(\phi
_1^t ,\theta
_1^t ,0)A\left( \lambda _{W^{+} } ,\lambda _b \right)
\eeq
where $\mu =\lambda _{W^{+} } -\lambda _b $ and $\lambda_1$
is
the $t$
helicity.  The final $W^{+}$ momentum is in the $\theta _1^t
,\phi _1^t$ direction, see Fig. 1.  For the $CP$-conjugate
process, $\bar t \rightarrow W^{-} \bar b$, in the $\bar t$
rest
frame
\beq
\langle \theta _2^t ,\phi _2^t ,\lambda _{W^{-} },\lambda
_{\bar b}|\frac 12,\lambda _2\rangle =D_{\lambda _2,\bar \mu
}^{(1/2)*}(\phi
_2^t ,\theta _2^t ,0)B\left( \lambda _{W^{-} },\lambda
_{\bar b}\right)
\eeq
with $\bar \mu =\lambda _{W^{-}}-\lambda _{\bar b}$,
$\lambda_2$ is
the $\bar t$
helicity.  So, by Lorentz invariance there are only 2
amplitudes $A(0, -1/2), A(-1, -1/2) $ for $b_L,$ etc.

Such formulas\footnote{To modularize the analysis, we
consistently use the standard helicity formalism and
phase conventions\cite{JW}.  Previous analyzes of the
helicity properties of $t \rightarrow W+ b$ include  [4,7];
studies of $CP$ violation in $t$ reactions include [4,8,9].
Input numbers: $m_t = 175GeV, m_W = 80.35GeV, m_b = 4.5GeV$.}
only assume
Lorentz invariance and do not
assume any
discrete symmetry properties.  Therefore, it is easy to
test for the consequences of addtional symmetries[6].
As shown in Table 1 a specific discrete symmetry implies a
specific
relation among the associated
helicity amplitudes.  In particular[6,10], measurement of a
non-real helicity
amplitude (i.e. of a non-zero relative phase $\beta_i$)
implies
a violation of $\tilde T_{FS}$ invariance when a first-order
perturbation in an
``effective" hermitian Hamiltonian is reliable.  $\tilde
T_{FS}$
invariance is
to be distinguished from canonical $T$ invariance which
requires
interchanging
``final'' and ``initial'' states. So $\tilde
T_{FS}$
invariance will be violated when either there is a violation
of
canonical $T$ invariance or when there are absorptive final-
state
interactions. Actual time-reversed
reactions are
required for a direct test of $T$ invariance.

Similarly in the $W^{+}$ rest frame, the matrix
element for $W^{+} \rightarrow l^+ \nu$ or for $W^{+}
\rightarrow j_{\bar d} j_{u}$ is
\beq
\langle \tilde \theta _a ,\tilde \phi _a ,\lambda _{l^+}
,\lambda
_{\nu}
| 1,\lambda _{W^+} \rangle =D_{\lambda _{W^+},1 }^{1*}
 (\tilde \phi
_a ,\tilde \theta
_a ,0)c
\eeq
since $\lambda
_{\nu}= - \frac{1}{2}, \lambda
_{l^+}= \frac{1}{2}$, neglecting $(\frac{m_l}{m_W})$
corrections.  This equation also describes the $W^{+}
\rightarrow j_{\bar d}
j_{u}$ decay mode, neglecting $(\frac{ m_{jet} }{m_W})$
corrections.

The associated composite decay density matrix for
$t\rightarrow
W^{+}b\rightarrow (l^{+}\nu )b$, or for $t\rightarrow
W^{+}b\rightarrow (j_{\bar
d}j_u)b$, is
\begin{equation}
R_{\lambda _1\lambda _1^{^{\prime }}}=\sum_{\lambda
_W,\lambda
_W^{^{\prime
}}}\rho _{\lambda _1\lambda _1^{^{\prime }};\lambda _W\lambda
_W^{^{\prime
}}}(t\rightarrow W^{+}b)\rho _{\lambda _W\lambda _W^{^{\prime
}}}(W^{+}\rightarrow l^{+}\nu )
\end{equation}
where $\lambda
_W,\lambda
_W^{^{\prime
}}=0, \pm 1 $ with

$$
\rho _{\lambda _1\lambda _1^{^{\prime }};\lambda _W\lambda
_W^{^{\prime
}}}(t\rightarrow W^{+}b)=\sum_{\lambda _b=\mp 1/2}D_{\lambda
_1,\mu
}^{(1/2)*}(\phi _1^t,\theta _1^t,0)D_{\lambda _1^{^{\prime
}},\mu ^{^{\prime
}}}^{(1/2)}(\phi _1^t,\theta _1^t,0)A(\lambda _W,\lambda
_b)A(\lambda
_W^{^{\prime
}},\lambda _b)^{*}
$$
$$
\rho _{\lambda _W\lambda _W^{^{\prime }}}(W^{+}\rightarrow
l^{+}\nu
)=D_{\lambda _W,1}^{1*}(\widetilde{\phi _a},\widetilde{\theta
_a}%
,0)D_{\lambda _W^{^{\prime }},1}^1(\widetilde{\phi
_a},\widetilde{\theta _a}%
,0) |c|^2
$$

This composite decay density matrix can be
elegantly
expressed in terms of eight helicity parameters $(\xi,
\sigma, \zeta,
 \ldots )$:
\ber
{\bf R=}\left(
\begin{array}{cc}
{\bf R}_{++} & e^{\iota \phi _1^t }
{\bf r}_{+-} \\ e^{-\iota \phi _1^t }{\bf r}_{-+} & {\bf
R}_{--}
\end{array}
\right)
\eer
The diagonal
elements  are
\beq
{\bf R}_{\pm \pm }={\bf n}_a[1\pm {\bf f}_a\cos \theta
_1^t ]\pm (1/\sqrt{%
2})\sin \theta _1^t \{\sin 2 \tilde \theta _a\
[\omega \cos \tilde \phi
_a+\eta ^{\prime }\sin \tilde \phi _a] - 2 \sin \tilde \theta
_a\
[\eta \cos \tilde \phi
_a+\omega ^{\prime }\sin \tilde \phi _a] \}
\eeq
In (5), ${\bf R}_{\pm \pm }$ are simply the angular
distributions $
\frac{dN}{d(\cos
\theta
_1^t )d(\cos \tilde \theta _a)d \tilde \phi_a} $ for the
polarized $t$ decay
chain, $t\rightarrow
W^{+}b\rightarrow (l^{+}\nu )b$, or $t\rightarrow
W^{+}b\rightarrow (j_{\bar
d}j_u)b$.  The off-diagonal elements depend on
\ber
\begin{array}{c}
{\bf r}_{+-}=({\bf r}_{-+})^{*} \\ ={\bf n}_a{\bf f}_a\sin
\theta _1^t
+ \sqrt{2} \sin \tilde \theta _a \{ \cos
\theta _1^t [\eta \cos \tilde
\phi _a+\omega ^{\prime }\sin \tilde \phi _a]+\iota [\eta
\sin \tilde \phi
_a-\omega ^{\prime }\cos \tilde \phi _a]\}\\
- \frac{1}{\sqrt{2} } \sin 2 \tilde \theta _a \{ \cos
\theta _1^t [\omega \cos \tilde
\phi _a+\eta ^{\prime }\sin \tilde \phi _a]+\iota [\omega
\sin \tilde \phi
_a-\eta ^{\prime }\cos \tilde \phi _a]\}
\end{array}
\eer
In (6,7),%
\ber
\begin{array}{c}
{\bf n}_a=\frac 1{8}(5-\cos 2\tilde \theta _a-\sigma
[1+3\cos 2\tilde
\theta _a]-4 [\xi-\zeta] \cos \tilde \theta _a ) \\
{\bf n}_a{\bf f}_a=\frac 1{8}(4 [1-\sigma] \cos \tilde \theta
_a -\xi [1+3\cos
2\tilde
\theta _a]+\zeta [5-\cos 2\tilde \theta _a])
\end{array}
\eer
or equivalently%
\ber
\left(
\begin{array}{c}
{\bf n}_a \\ {\bf n}_a{\bf f}_a
\end{array}
\right) =\sin ^2\tilde \theta _a\frac{\Gamma _L^{\pm
}}{\Gamma }\pm
\frac 14(3+\cos 2\tilde \theta _a ) \frac{\Gamma _T^{\pm
}}{\Gamma
}\mp \cos \tilde \theta _a  \frac{\Gamma _T^{\mp }}{\Gamma
}
\eer
The subscripts on
the $\Gamma $'s denote the
polarization of the final $W^{+}$, either
``L=longitudinal'' or
``T=transverse''; superscripts denote ``$\pm $ for
sum/difference of
the $b_{L\ }$versus $b_R$ contributions''.

Similarly, for the CP-conjugate process  $\bar t \rightarrow
W^{-} \bar b \rightarrow (l^{-} \bar\nu ) \bar b$ or $\bar t
\rightarrow W^{-} \bar b \rightarrow (j_{d}j_{\bar u}) \bar
b$
\ber
{\bf \bar R=}\left(
\begin{array}{cc}
{\bf \bar R}_{++} & e^{\iota \phi _2^t }
{\bf \bar r}_{+-} \\ e^{-\iota \phi _2^t }{\bf \bar r}_{-
+} & {\bf \bar R}%
_{--}
\end{array}
\right)
\eer
\beq
{\bf \bar R}_{\pm \pm }={\bf n}_b[1\mp {\bf f}_b\cos \theta
_2^t ]\mp (1/%
\sqrt{2})\sin \theta _2^t  \{\sin 2\tilde \theta _b [ \bar
\omega \cos \tilde
\phi _b-\bar \eta ^{\prime }\sin \tilde \phi _b] -2 \sin
\tilde
\theta _b [ \bar \eta \cos \tilde
\phi _b-\bar \omega ^{\prime }\sin \tilde \phi _b] \}
\eeq
\ber
\begin{array}{c}
{\bf \bar r}_{+-}=({\bf \bar r}_{-+})^{*} \\ =-{\bf n}_b{\bf
f}_b\sin \theta
_2^t - \sqrt{2} \sin \tilde \theta _b \{\cos \theta _2^t
[\bar
\eta \cos \tilde \phi _b-\bar \omega ^{\prime }\sin \tilde
\phi _b]+\iota
[\bar \eta \sin \tilde \phi _b+\bar \omega ^{\prime } \cos
\tilde \phi _b]\} \\
+ \frac{1}{\sqrt{2} }  \sin 2 \tilde \theta _b \{\cos \theta
_2^t [\bar
\omega \cos \tilde \phi _b-\bar \eta ^{\prime }\sin \tilde
\phi _b]+\iota
[\bar \omega \sin \tilde \phi _b+\bar \eta ^{\prime } \cos
\tilde \phi _b]\}
\end{array}
\eer
\ber
\begin{array}{c}
{\bf n}_b=\frac 1{8}(5-\cos 2\tilde \theta _b-\bar \sigma
[1+3\cos
2\tilde \theta _b]-4[\bar \xi - \bar \zeta]\cos
\tilde \theta _b) \\
{\bf n}_b{\bf f}_b=\frac 1{8}(4[1 - \bar \sigma]\cos
\tilde \theta _b -\bar \xi
[1+3\cos 2\tilde \theta _b]+\bar \zeta [5-\cos
2\tilde \theta _b])
\end{array}
\eer
\ber
\left(
\begin{array}{c}
{\bf n}_b \\ {\bf n}_b{\bf f}_b
\end{array}
\right) =\sin ^2\tilde \theta _b\frac{\bar \Gamma _L^{\pm
}}{\bar \Gamma }%
\pm \frac 14(3+\cos 2\tilde \theta _b) \frac{\bar \Gamma
_T^{\pm
}}{\bar
\Gamma } \mp \cos \tilde \theta _b \frac{\bar \Gamma _T^{\mp
}}{\bar
\Gamma }
\eer

\subsection*{Definitions of helicity parameters by
partial-width
intensities \newline for
polarized-final-states:}

There are eight $t \rightarrow W^+ b$ decay parameters since
there are the four $W_{L,T}$ $b_{L,R}$ final-state
combinations:
The first parameter is simply $\Gamma \equiv
\Gamma
_L^{+}+\Gamma _T^{+}$, i.e. the partial width
for $t\rightarrow W^{+} b$, and
\begin{equation}
\begin{array}{c}
\Gamma _L^{\pm }=\left| A(0,-\frac 12)\right| ^2\pm \left|
A(0,\frac
12)\right| ^2 \\
\Gamma _T^{\pm }=\left| A(-1,-\frac 12)\right| ^2\pm \left|
A(1,\frac
12)\right| ^2.
\end{array}
\end{equation}
The second is the $b$ quark's
chirality
parameter $\xi \equiv \frac 1\Gamma (\Gamma _L^{-}+\Gamma
_T^{-})$.
Equivalently, \newline  \hspace{1pc} $  \xi \equiv$ (Prob
$b$ is
$b_L$) $ - $
(Prob $b$ is $b_R$),
\begin{equation}
\xi \equiv |< b_L | b >|^{2} - |< b_R
| b
>|^{2}
\end{equation}
So a value $\xi = 1$ means the coupled $b$ quark is
pure $b_L$, i.e. $\lambda_b = -1/2$.   The remaining two
partial-width parameters are defined by
\begin{equation}
\zeta \equiv (\Gamma _L^{-}-\Gamma _T^{-})/ \Gamma ,
\hspace{1pc} \sigma \equiv (\Gamma
_L^{+}-\Gamma
_T^{+})/ \Gamma .
\end{equation}
This implies for $W^+$ polarimetry that  \newline  $
\hspace{2pc}
\sigma
=$ (Prob
$W^{+}$ is $W_L$)
$
-
$ (Prob $W^{+}$ is $W_T$), \newline is the analogue of
the
$b$ quark's chirality parameter in
(16).
Thus, the parameter $\sigma$ measures
the
degree of polarization, ``L minus T", of the emitted $W^+$.
For a pure $(V - A)$, or $(V + A)$, coupling $\sigma = 0.41$,
see Table 1.

The interference between these $W_L$ and
$W_R$ amplitudes can also be determined by measuring the
four normalized parameters,
\begin{equation}
\begin{array}{c}
\omega \equiv I_{
{\cal R}}^{-}\ / \Gamma , \hspace{2pc}  \eta
\equiv I_{
{\cal R}}^{+}\ / \Gamma  \\
\omega ^{\prime
}\equiv I_{
{\cal I}}^{-}\ / \Gamma , \hspace{2pc} \eta
^{\prime }\equiv
I_{{\cal I}%
}^{+}\ / \Gamma
\end{array}
\end{equation}
The associated $W_L - W_T$ interference intensities are
\begin{equation}
\begin{array}{c}
I_{{\cal R}}^{\pm }=\left| A(0,-\frac 12)\right| \left| A(-
1,-\frac
12)\right| \cos \beta _a  \cr \pm \left| A(0,\frac
12)\right|
\left| A(1,\frac
12)\right| \cos \beta _a^R  \\
I_{{\cal I}}^{\pm }=\left| A(0,-\frac 12)\right| \left| A(-
1,-\frac
12)\right| \sin \beta _a \cr \pm \left| A(0,\frac
12)\right|
\left| A(1,\frac
12)\right| \sin \beta _a^R
\end{array}
\end{equation}
Here,  $\beta _a\equiv
\phi _{-1}^a-\phi _0^a$, and $\beta
_a^R\equiv \phi
_1^a-\phi _0^{aR}$\ are the measurable phase differences of
of the
associated helicity amplitudes
$A(\lambda_{W^+},\lambda_b)=\left|
A\right| \exp \iota \phi $ in the standard helicity amplitude
phase convention\cite{JW}.

When there is only a $(V-A)$
coupling and $m_b = 0$ these parameters have the
values\footnote{If one factors out ``W-polarimetry factors"
,see footnote below, via $ \sigma = {\cal S}_W \tilde
\sigma$,
$\omega = {\cal R}_W \tilde \omega$, $\ldots$ these
parameters all equal one or zero for pure $(V-A)$ and $m_b
=0$.  Systematic effects will cancel by considering the
ratios $\zeta
/ \xi$ versus ${\cal S}_W $, and $\omega / \xi$ versus ${\cal
R}_W $
in the two pre-SSB tests.}
shown in Table
2.  Note that the ``pre-SSB case" of a
mixture of
only $V$ and $A$
couplings and $m_b = 0 $  implies that the two parameters
directly sensitive to $
\tilde{T}_{FS} $ violation vanish, ${\omega ^{\prime } } =
{\eta ^{\prime }
} = 0$.  Also in the pre-SSB case, the $b$ quark's chirality
parameter $\xi \rightarrow \frac{\left|
g_L\right|
^2-\left|
g_R\right| ^2}{\left| g_L\right| ^2+\left|g_R\right| ^2}$ so
that the ``stage-one spin correlation" parameter
$\frac{\zeta}{ {\cal S}_W }
\rightarrow \xi$.  So, in this special case $\zeta$
also measures
the $b$ quark's helicity and  $\frac{\zeta}{ {\cal S}_W } =
\xi$, but for
more general
couplings and/or $m_b \neq 0 $
neither is true. Also in the pre-SSB case, the interference
parameter $\frac{\omega}{ {\cal R}_W } =
\xi$.  Therefore, precision measurements with $\xi$ and
$\zeta$ distinct, and with $\xi$ and $\omega$ distinct, will
be two useful probes of the dynamics
of EW spontaneous symmetry breaking, see (26) and (27) below
for instance.

From Table 2, one easily sees
that the numerical values of ``$\xi, \zeta, \sigma, \ldots $"
are very different for unique Lorentz couplings. This is
indicative of the analyzing power of polarization techniques
in two-body $t$ decay modes. Both the real and
the imaginary parts of the associated helicity amplitudes can
be directly measured, c.f. (19).

\subsection*{Tests for anomalous $\Gamma_L$, $\Gamma_T$
polarized-partial-widths:}

The contribution of the longitudinal($L$) and transverse($T$)
$W$-amplitudes in the decay
process is projected out by the simple formulas:
\begin{eqnarray*}
I_{
{\cal R}}^{b_L,b_R}\equiv \frac 12(I_{{\cal R}}^{+}\pm
I_{{\cal R}%
}^{-})=|A(0,\mp \frac 12)||A(\mp 1,\mp \frac 12)|\cos \beta
_a^{L,R} =\frac
\Gamma 2({\eta} \pm {\omega} )
\end{eqnarray*}
\begin{eqnarray*}
I_{
{\cal I}}^{b_L,b_R}\equiv \frac 12(I_{{\cal I}}^{+}\pm
I_{{\cal I}%
}^{-})=|A(0,\mp \frac 12)||A(\mp 1,\mp \frac 12)|\sin \beta
_a^{L,R} =\frac
\Gamma 2({\eta^{\prime }} \pm {\omega^{\prime }}
)
\end{eqnarray*}
\begin{eqnarray*}
\Gamma _L^{b_L,b_R}\equiv \frac 12(I_L^{+}\pm I_L^{-
})=|A(0,\mp \frac
12)|^2
=\frac \Gamma 4(1+{\sigma} \pm \xi \pm {\zeta} )
\end{eqnarray*}
\begin{eqnarray}
\Gamma _T^{b_L,b_R}\equiv \frac 12(I_T^{+}\pm I_T^{-
})=|A(\mp 1,\mp
\frac 12)|^2
=\frac \Gamma 4(1-{\sigma} \pm \xi \mp {\zeta} )
\end{eqnarray}
In the first line, $\beta _a^L=\beta _a$. Unitarity, requires
the two right-triangle
relations
\ber
(I_{
{\cal R}}^{b_L})^2+(I_{{\cal I}}^{b_L})^2=\Gamma
_L^{b_L}\Gamma
_T^{b_L} \\
(I_{{\cal R}}^{b_R})^2+(I_{{\cal I}}^{b_R})^2=\Gamma
_L^{b_R}\Gamma _T^{b_R}.
\eer

It is important to determine directly from experiment
whether or not these partial widths are anomalous in nature
versus the standard $(V-A)$ predictions
because the
$W_L$ and $W_T$ partial widths might have distinct
dynamical differences versus the SM predictions if
electroweak dynamical symmetry
breaking(DSB) occurs in nature, e.g. associated with a $t$
quark compositeness and/or a $(t
\bar t)$ condensate and/or anomalous $W_{L,T}$-$W_{L,T}$
interactions.  In areas of physics in which DSB does
occur, the on-shell and off-mass-shell values of such
polarized-partial-widths reveal important dynamical
information.

\subsection*{Tests for non-CKM-type CP, and $
\tilde{T}_{FS} $
violations:}

A violation of $\tilde T_{FS}$-invariance could
occur for a dynamical reason, e.g. because of the exchange of
an unknown Z$%
^{^{\prime }}$ boson between the final $W^{+}$and
the final $%
b$ in which the Z$^{^{\prime }}$ couples differently
to
the $W_L$
versus the $W_T$. Or $\tilde T_{FS}$-violation
could occur
because of a fundamental violation of canonical
$T$-invariance. Whatever the
cause might turn out to be, the experimental discovery of a
violation of $%
\tilde T_{FS}$-invariance in $t \rightarrow W^+ b$ would
be very significant.
\begin{itemize}
    \item If the primed parameters $ \omega ^{\prime } \neq 0
$ and/or $
\eta
^{\prime }
\neq 0
\Longrightarrow  $ {\bf $\tilde{T}_{FS} $ is violated:}
\end{itemize}
Only two of the four parameters $\eta ,\eta
^{^{\prime }},\omega ,\omega ^{^{\prime
}}$ are needed to test for $ \tilde{T}_{FS} $
violation because the two-right-triangle relations imply
\begin{equation}
({\eta} \pm {\omega} )^2+({\eta^{\prime }}
\pm
{\omega^{\prime }} )^2=\frac
14[(1 \pm \xi )^2-( {\sigma} \pm {\zeta} )^2].
\end{equation}

The barred parameters $ \bar{\xi},
\bar{\zeta}, \ldots $ have
the analogous
definitions for the CP conjugate modes, for instance,
$ \bar \xi =$ (Prob $\bar b $
is $\bar b_R$)
$ - $ (Prob $\bar b $ is $\bar b_L$) ,
\begin{equation}
\begin{array}{c}
 \hspace{2pc} \bar \Gamma _L^{\pm }=|B(0,\frac 12)|^2\pm
|B(0,-\frac
12)|^2\\
\bar I_{{\cal R}}^{\pm }=\left| B(0,\frac 12)\right| \left|
B(
1,\frac
12)\right| \cos \beta _b \pm \left| B(0,-\frac
12)\right|
\left| B(-1,-\frac
12)\right| \cos \beta _b^L.
\end{array}
\end{equation}
Therefore,
\begin{itemize}
    \item If any $
\bar{\xi} \neq \xi,
\bar{\zeta} \neq \zeta, \ldots $ $ \Longrightarrow $ {\bf CP
is
violated:}
\end{itemize}
With precisions forseeable in the near future, since $m_b/m_t
\sim m_b/m_W \sim 0$, there are two simple
tests for ``non-CKM-type" CP violation in $t
\rightarrow W b $ decay. The two tests are: $
\beta _a=\beta _b $, where $\beta _a=\phi _{-1}^a-\phi _0^a$,
$\beta _b=\phi _1^b-
\phi
_0^b$, and
$
r_a=r_b $, where $
r_a=\frac{|A\left( -1,-\frac 12\right) |}{|A\left( 0,-\frac
12\right)
|}$, $ r_b=\frac{|B\left( 1,\frac 12\right) |}{|B\left(
0,\frac
12\right) |} $.   Normally[10] a CKM-phase
will contribute equally at tree level to both the $t
\rightarrow W^+ b_L$
decay amplitudes and so will cancel out in the ratio of their
moduli and in their relative phase.  With improved precision
there are the additional tests:  $\beta_a^R=\beta_b^L$,
$r_a^R = r_b^L$, and $\lambda_R = \bar \lambda_L$, see
[6,10].

\subsection*{Tests for additional Lorentz
structures:}

A chiral classification of
additional Lorentz structure is a natural phenomenological
extension
of the
symmetries of the SM, but this formalism easily allows
searches for
non-chiral couplings[6,10].
For \hskip1em  $t \rightarrow W^+ b$, the most general
Lorentz
coupling is
\begin{equation}
W_\mu ^{*}\bar u_{b}\left( p\right) \Gamma ^\mu
u_t \left(
k\right)
\end{equation}
where $k_t =q_W +p_b $. In (25)
\begin{eqnarray*}
\Gamma _V^\mu =g_V\gamma ^\mu +
\frac{f_M}{2\Lambda }\iota \sigma ^{\mu \nu }(k-p)_\nu   +
\frac{g_{S^{-}}}{2\Lambda }(k-p)^\mu  \\
+\frac{g_S}{2\Lambda
}(k+p)^\mu
+%
\frac{g_{T^{+}}}{2\Lambda }\iota \sigma ^{\mu \nu }(k+p)_\nu
\end{eqnarray*}
\begin{eqnarray*}
\Gamma _A^\mu =g_A\gamma ^\mu \gamma _5+
\frac{f_E}{2\Lambda }\iota \sigma ^{\mu \nu }(k-p)_\nu \gamma
_5
+
\frac{g_{P^{-}}}{2\Lambda }(k-p)^\mu \gamma
_5  \\
+\frac{g_P}{2\Lambda }%
(k+p)^\mu \gamma _5  +\frac{g_{T_5^{+}}}{2\Lambda }\iota
\sigma ^{\mu \nu
}(k+p)_\nu \gamma _5
\end{eqnarray*}
The parameter
$%
\Lambda =$ ``the effective-mass scale of new physics''. In
effective field
theory
this
is the scale at which new particle thresholds are expected to
occur or where the theory becomes non-perturbatively
strongly-interacting so as to overcome perturbative
inconsistencies.  It can also be interpreted as a measure of
a
new compositeness scale.  In old-fashioned renormalization
theory
$\Lambda$  is the scale at
which the calculational methods and/or the principles of
``renormalization''
breakdown.

Without additional theoretical or
experimental
inputs, it is not possible to select what is the ``best"
minimal set of couplings for
analyzing the structure of the $(\bar b t)$ charged current.
For
instance, there are the
equivalence theorems that for the
vector current, $
S\approx V+f_M , T^{+}\approx -V+S^{-}$,
and for the axial-vector current,
$
P\approx -A+f_E , T_5^{+}\approx A+P^{-}$.
So, from the perspective of searching for fundamental
dynamics, it
is important to investigate what limits can be set for a
variety of Lorentz structures (including $S^{\pm}$,
$P^{\pm}$, $T^{\pm}$, and ${T_5}^{\pm}$) and not just for a
kinematically minimal, but theoretically prejudiced, set.

DSB or compositeness considerations motivate
searching for an additional tensorial $g_{+}=f_M
+
f_E$ coupling which would preserve the three logically
independent $m_b = 0$ signatures for only $b_L$ couplings:
$\xi =1, \sigma = \zeta , \eta = \omega $, and also
$\omega^{\prime} = \eta^{\prime} $ if $ \tilde{T}_{FS} $
violation occurs. But $g_{+} $ would give
non-($V-A$)-values, $\sigma =\zeta
\neq 0.41$ and $%
\eta =\omega \neq 0.46$. For $\Lambda$ large, there are the
predictions for purely real $g_+$ that\footnote{In the
presence of additional Lorentz structures,
W-polarimetry factors
${\cal S}_W =\frac{1-2\frac{m_W
^2}{m_t ^2}}{1+2\frac{m_W ^2}{m_t ^2}}= 0.4068 $, and $
{\cal R}_W =\frac{\sqrt{2}\frac{m_W}m_t}{1+2\frac{m_W
^2}{m_t ^2}}= 0.4567 $
naturally appear because of the referencing of
``new physics" to the $(V-A)$ structure of the SM, see
[6].  See also, footnote 3 above.}
\begin{equation}
(\frac{\zeta}{ {\cal S}_W } - \xi)=(\xi - \frac{\omega}{
{\cal R}_W
} )\frac gl = 0.076 (\xi - \frac{\omega}{ {\cal R}_W
} )
\end{equation}
and for purely imaginary $g_+$ that
\begin{equation}
(\frac{\zeta}{ {\cal S}_W } - \xi )=( \xi - \frac{\omega}{
{\cal R}_W
} )\frac{2u}{o} = 8.4 (\xi - \frac{\omega}{ {\cal R}_W
} )
\end{equation}
The ratios ``$g/l$" and $2u/o$" are given functions [6] of
$m_{t,W}$.

While some terms of non-(V-A) form do occur as higher-order
perturbative-corrections
in
the standard model,
such SM
contributions  are expected to be less than the precision of
planned Tevatron and LHC experiments. Additional systematic
analysis of the
higher-order QCD and EW corrections versus the tree-level (V-
A) predictions for and the sensitivities of S2SC
parameters $( \xi, \sigma, \ldots; r_a, \beta_a, \ldots ;
\Gamma _L^{\pm
}, \ldots )$ should be performed, c.f. Jezabek and Kuhn [7].
The $S^{-}\pm P^{-}$
couplings do not contribute when the final $W^+$ is on the
mass-shell.
Both the weak magnetism  $\frac{f_M}{2\Lambda }$ and the weak
electricty $%
\frac{f_E}{2\Lambda }$ terms are divergenceless. On the other
hand, since $%
q^2={m_W}^2$,  even when $m_b =m_t $ the couplings $S^{-
},T^{+},A,P^{-},T_5^{+}$ contribute to the divergences.

To study the reality structure of  ${J_{\bar b t}}$, we
assume[11,10] that it is Hermitian
and has an SU(2) symmetry $t \leftrightarrow b $.
Then for
real form factors, the
``Class I'' couplings are $V,A,f_M,P^{-}$, and
``Class
II'' couplings are  $f_E,S^{-}$.  We define  $J_{\bar b
t}^\mu
=$
$J_I^\mu +$ $%
J_{II}^\mu $ where for $U=\exp (\iota \pi I_2)$,
$$
\begin{array}{cc}
(J_I^\mu )^{\dagger }=-UJ_I^\mu U^{-1} & {\textstyle First
Class }  \\
(J_{II}^\mu )^{\dagger }=UJ_{II}^\mu U^{-1} & {\textstyle
Second Class }
\end{array}
$$
There
is a ``clash''
between the ``Class I and Class II'' structures and
the consequences of
time-reversal invariance. Useful
theorems are
that (a) ($t \leftrightarrow b $ symmetry) + ($T$
invariance) $%
\Longrightarrow $ Class II currents are absent, (b) ($t
\leftrightarrow
b $ symmetry) + (existence of $J_I^\mu $ and
$J_{II}^\mu
$) $%
\Longrightarrow $ violation of $T$ invariance, and (c)
(existence
of
$%
J_{II}^\mu $) +  ($T$ invariance) $\Longrightarrow $($t
\leftrightarrow b $ symmetry) in $J_{\bar b t} $ is
broken.

\subsection*{Measurement of relative phase of $b_L$ and $b_R$
amplitudes:}

If only $b_L$ coupling's existed, there would be only 2
amplitudes, so 3 measurements,  of $r_a,\beta _a,$and the
partial width $\Gamma$
would provide
a ``complete measurement'' of $t \rightarrow
W^+ b $. In SM, these respectively equal $0.650, 0, \sim 1.55
GeV$. However, since $m_b \neq 0$,
there are 2 more amplitudes, $A(0,\frac 12)$ and $A(1,\frac
12)$, so to
achieve
an ``almost'' complete measurement, 3 additional quantities
must
be
determined, e.g.   $r_a^R \equiv \frac{|A(1,\frac
12)|}{|A(0,\frac 12)|} ,\beta _a^R$
and $%
\lambda _R\equiv \frac{|A(0,\frac 12)|}{|A(0,-\frac 12)|}$.
In SM, these respectively equal $3.08, 0, 0.0069$.
However, a further measurement is necessary to determine the
relative phase of the $b_L$ and $b_R$ amplitudes,
$\beta
_a^o\equiv
\phi _o^{aR}-\phi _o^a$ or $\beta _a^1\equiv \phi _1^a-\phi
_{-
1}^a$. This would be possible by development of $b$
quark-polarimetry techniques.  It might be possible to obtain
this
missing phase through a measurement of the interference
between the $b_L$ and $b_R$ amplitudes in $t \rightarrow  W^+
b $ with $b \rightarrow c l^- \bar \nu$ where the b quark is
required to fragment in a $J \neq 0$ mass
region, e.g. $B^{**}, \Lambda_b, \ldots $.
Existing
LEP and SLD data on the associated jet-lepton angular
distributions could
be
analyzed to see if $b$ quark-polarimetry is feasible in $t$
decays.

\subsection*{Sensitivities of measurements at hadronic
colliders:}

The simplest kinematic measurement of the above helicity
parameters\footnote{Their cleanest measurement would
presumably be at a future $e^- e^+$ or $\mu^- \mu^+$
collider.} at the Tevatron and at the LHC would be through
purely
hadronic top decay modes.  CDF has reported[12]
observation of
such decays.  In this case the $(t \bar t)_{cm}$ frame is
accessible and the four-variable, stage-two spin-correlation
function, $I_4$,
\ber
\textstyle{I(E}_{W^+} \textstyle{,E}_{W^-}\textstyle{,}\tilde
\theta
_1\textstyle{,}\tilde
\theta _{2\textstyle{ }}\textstyle{) = } \sum_{i} \{ \rho_{+-
} (q_i
\bar q_i
\rightarrow t \bar t )^{\textstyle{prod}} [ \rho
_{++}\bar
\rho _{--} + \rho
_{--}\bar
\rho _{++} ]  \nonumber  \\ + \rho_{++} (q_i \bar q_i
\rightarrow
t
\bar t ) ^{\textstyle{prod}} [\rho
_{++}\bar \rho
_{++}
 +\rho _{--}\bar \rho _{--} ] \}
\eer
can be used.  The ``$i$" summation is over the incident
quarks and gluons in the $p \bar p$, or $pp$. If one thinks
in terms of
probabilities, the quantum-mechanical structure of this
expression is obvious.
The kinematic variables in $I_4$ are the $(t \bar t)$
center-of-mass energies $E_{W^+}$ and $E_{W^-}$, the polar
angle
$\tilde \theta _1$ which specifies the $j_{\bar d}$ jet (or
the
$l^{+}$)
momentum in the $W^{+}
$rest frame when the boost is directly from the $(t \bar
t)_{cm}$ frame\cite{PR}, and $\tilde \theta _2$ for the
$j_{d}$ jet
(or the
$l^{-}$)
momentum in the $W^{-}
$rest frame.

Such ``$S2SC$" functions may turn out to be very useful for
probing for unexpected $t \bar t$ production mechanisms
since,
by CP invariance, $I_4$ only depends on two diagonal
production-density-matrix-elements\cite{PR} in the $q
\bar q \rightarrow t \bar t $ channel and also in the $ g g
\rightarrow t
\bar
t $ channel.  CP can also be tested in production [4,8,6].

In (28), the composite decay density matrix elements are
simply
the
decay
probability for a $t_1$ with
helicity $\frac h2$ to decay $t \rightarrow W^{+} b$ followed
by  $W^{+}\rightarrow j_{\bar
d}j_u$, or $%
W^{+}\rightarrow l^{+}\nu $  since
\newline
${d\textstyle{N}}/{d\left( \cos \theta _1^t \right) d\left(
\cos
\tilde
\theta _1\right) }=\rho _{hh}\left( \theta _1^t ,\tilde
\theta
_1 \right)  $ and for the decay of the $\bar t_2$ , $\bar
\rho
_{hh}=
\newline
\rho _{-h,-
h} \left(  1
\rightarrow 2, \textstyle{add bars} \right)$.
For $t_1$ with
helicity $\frac h2$
\beq
\rho _{hh}= \rho_o+h \rho_c \cos \theta _1^t +h \rho_s \sin
\theta _1^t
\eeq
where
\ber
\rho_o =\frac 18 \{6- 2 \cos ^2\omega
_1\cos^2\tilde \theta
_1- \sin ^2\omega _1\sin ^2\tilde \theta _1
\nonumber \\
+\sigma [ 2- 6 \cos ^2\omega
_1\cos^2\tilde \theta
_1- 3 \sin ^2\omega _1\sin ^2\tilde \theta _1 ]
-4(\xi-\zeta) \cos \omega
_1 \cos \tilde \theta
_1 \}\\
\rho_c =\frac 18 \{\zeta [6- 2 \cos ^2\omega
_1\cos^2\tilde \theta
_1- \sin ^2\omega _1\sin ^2\tilde \theta _1]
\nonumber \\
+\xi [ 2- 6 \cos ^2\omega
_1\cos^2\tilde \theta
_1- 3 \sin ^2\omega _1\sin ^2\tilde \theta _1 ]
+4(1-\sigma) \cos \omega
_1 \cos \tilde \theta
_1 \}\\
\rho_s =\frac{1}{\sqrt{2}} \{\frac{1}{2} \omega \sin 2\omega
_1 [ \sin ^2\tilde \theta _1 -  2 \cos^2\tilde \theta
_1]
+2 \eta \sin \omega
_1 \cos \tilde \theta
_1 \}
\eer
with\footnote{The rotation by
$\omega_1$ about the implicit $y_a$ axis in Fig. 1 is given
by
$ \sin \omega_1 = m_W \beta \gamma \sin \theta_1^t / p_1$, $
\cos \omega_1 = \frac{E_{cm} (m_t^2 - m_W^2 + [ m_t^2 +
m_W^2]
\beta \cos \theta_1^t ) }{4m_t^2 p_1} $ where $p_1=$ the
magnitude of the $W^+$ momentum in the $(t \bar t)_{cm}$
frame
and $\gamma,\beta$ describe the boost from this $cm$ frame to
the $t_1$ rest frame [$\gamma= E_{cm}/(2m_t)$ with $E_{cm}=$
total energy of $t \bar t$, in $(t \bar t)_{cm}$].  } the
Wigner
rotation angle $\omega_1 =
\omega_1(E_{W^+})$.  Note that the $\rho_s$ term depends only
on the $W_L - W_T$ interference intensities, whereas the
$\rho_o$ and $\rho_c$ terms only depend on the polarized-
partial-widths, specifically
\ber
\rho_{o,c} =\frac 12 [ 2- 2 \cos ^2\omega
_1\cos^2\tilde \theta
_1- \sin ^2\omega _1\sin ^2\tilde \theta _1 ]
\frac{\Gamma_L^{\pm} } {\Gamma}
\nonumber \\
\pm \frac 14 [ 2+ 2 \cos ^2\omega
_1\cos^2\tilde \theta
_1+ \sin ^2\omega _1\sin ^2\tilde \theta _1 ]
\frac{\Gamma_T^{\pm} } {\Gamma}
\mp \cos \omega
_1 \cos \tilde \theta
_1 \frac{\Gamma_T^{\mp} } {\Gamma}
\eer
with $\bar \rho_{o,c} = \rho_{o,c}$  $ \left( 1 \rightarrow
2,
\textstyle{add bars} \right)$.

$I_4$ can also be written in terms of the variables useful
for
testing for non-CKM-type CP violation[the overall factor
$\Gamma_L^+
/ \Gamma$ is suppressed] :
\ber
\rho_{o,c} =\frac 12 [ 2- 2 \cos ^2\omega
_1\cos^2\tilde \theta
_1- \sin ^2\omega _1\sin ^2\tilde \theta _1 (1\pm
\lambda_R^2 ) ]
\nonumber \\
\pm \frac 14 [ 2+ 2 \cos ^2\omega
_1\cos^2\tilde \theta
_1+ \sin ^2\omega _1\sin ^2\tilde \theta _1 ] ( r_a^2\pm
[\lambda_R r_a^R ]^2 )
\mp \cos \omega
_1 \cos \tilde \theta
_1 ( r_a^2\mp [ \lambda_R r_a^R ]^2 ) \\
\rho_s =\frac{1}{\sqrt{2}} \{ \frac{1}{2} \sin 2\omega _1 [
\sin ^2\tilde \theta _1 -  2 \cos^2\tilde \theta
_1 ]  [r_a \cos \beta_a - \lambda_R r_a^R \cos \beta_a^R ]
\nonumber \\
+2 \sin \omega
_1 \cos \tilde \theta
_1 [r_a \cos \beta_a + \lambda_R r_a^R \cos \beta_a^R ] \}
\eer
with $\bar \rho_i = \rho_i ( 1 \rightarrow 2, a \rightarrow b
)$, $\lambda _R\equiv $ $\frac{|A\left( 0,\frac 12\right)
|}{|A\left(
0,-\frac 12\right) |}$ [6,10].

In the framework of the parton model, we characterize
the ``sensitivity" for measurement of a parameter ``a"
appearing in a spin-correlation function $I(x,y)=Z_o (x,y) +
a
Z_1 (x,y) +a^2 Z_2 (x,y) $  by the fractional uncertainty
``$\sigma_a / a $" where
\beq
\sigma_a \equiv \{ \sum_{ij} \frac{1}{\sigma_{ij} ^2 } [Z_1
(x_i,y_j) + 2a Z_2 (x_i,y_j) ]^2 \}^{-1/2}
\eeq
where $\sigma_{ij}=\surd I(x_i,y_j) $. ``$x,y$" are event
by event observables $E_{W^{+}},\tilde \theta_1, \ldots$. The
spin-correlation function $I(x_i,y_j)$ and the
quantity $[Z_1 (x_i,y_j) + 2a Z_2 (x_i,y_j)]$ are smeared
over
the parton distribution functions for the $p$ and $\bar p$
hadrons, just as
in
the evaluation of the $p \bar p \rightarrow t \bar t X $
cross
section,.
When ``a" only appears linearly, set $Z_2=0$.  This a simple
and natural sensitivity
criteria because for spin-correlation analyzes, a physical
consequence of the QM-factorization structure of the parton
model is that there are incident
parton longitudinal beams characterized by the Feynman $x_1$
and
$x_2$ momentum fractions instead of the known $p$ and $\bar
p$
momenta.  If the parton momenta were known so that this
momentum-smearing were not necessary, this procedure would
correspond to the usual ideal statistical error procedure for
characterizing least-squares measurements of
fundamental parameters[13].

Using this criteria, for measurements at the Tevatron and the
LHC we obtain the sensitivities listed in Table 3.  Since
these fractional uncertainties depend on $ \surd \frac{1}{ \#
events}$,
they can be scaled by the reader.  The numbers listed assume
$3 \cdot
10^4$
events in $p \bar p$ at $2 TeV$, $10^6$ events in $p p$ at
$14 TeV$, and the MRS($A^{'}$)
parton distribution set[14].

At the Tevatron, percent level uncertainties are typical for
measurements of
the helicity parameters $\xi, \zeta, \sigma, \omega, \eta$.
At the LHC, several mill level
uncertainties are typical.   These are also the sensitivity
levels found for measurement of the
polarized-partial-widths, $\Gamma_{L,T}^{\pm}$, and for the
non-CKM-type CP violation parameter $r_a$.
From $I_4$ the $\eta$
parameter($\omega$ parameter) can respectively best be
measured at the
Tevatron(LHC).  However, by use of additional variables (all
of $\tilde \theta_1, \tilde \phi_1, \tilde \theta_2, \tilde
\phi_2$) in the stage-two step of the decay sequences where
$W^{\pm} \rightarrow j_{\bar d,d} j_{u,\bar u}$, and/or
$l^{\pm}
\nu$,  we expect that these sensitivities
would then be comparable to that for the other helicity
parameters. Inclusion of additional variables should also
improve the sensitivity to the CP violation parameter
$\beta_a$ which is at $33^o$ (Tevatron), $9.4^o$ (LHC).
In regard
to effective mass-scales for new physics
exhibited by additional Lorentz couplings, we find
$50-70 TeV$ effective-mass scales can be probed at the
Tevatron and $110-750 TeV$ scales at the LHC. Associated
sensitivities for measurements via other spin-correlation
functions and with other top decay modes, in
particular when $W^{+} \rightarrow l^+ \nu$ and $W^{-}
\rightarrow l^- \bar \nu$, will be reported in a later paper.

We thank experimental and theoretical physicists for
discussions and assistance, in particular with respect to
matters specific to hadron colliders.  We thank Ming
Yang; and for computer services, John Hagan,
Christine Place-Sweet, and Mark Stephens. This work was
partially
supported by U.S. Dept. of Energy Contract No. DE-FG
02-96ER40291.

\newpage

\begin{center}
{\bf Table Captions}
\end{center}

Table 1: The helicity formalism is based on the assumption of
Lorentz invariance but not on any specific discrete symmetry
properites of the fundamental amplitudes, or couplings.  For
$%
t\rightarrow W^{+}b$ and $%
\bar t \rightarrow W^{-} \bar b$ a specific discrete
invariance
implies a definite symmetry relation among the associated
helicity amplitudes, such as shown.

Table 2: Comparison of helicity parameters' values for unique
Lorentz
couplings:  The first 3
parameters ($ \xi, \zeta, \sigma $), plus the partial
width $ \Gamma ( t\rightarrow W^{+}b ) $, give the
polarized-final-state partial-widths, $\bf{\Gamma}$, for the
four $W_{L,T} b_{L,R}$ final-state combinations.  The
latter 4 parameters ($ \omega, \eta, \omega^{'}, \eta^{'} $)
give the complete $W_L$ -$W_T$ interference
intensities, $\bf{\Gamma_{LT}}$. Numerical values are to two
digits for $m_b=0$.  Only at three digits, do the standard
model's $(V - A)$ values become sensitive to the $m_b$ mass;
see SM column in next table.

Table 3: At 2 TeV and at 14 TeV:  First, sensitivities versus
standard
model values for measurements of the fundamental parameters $
\xi, \zeta,
\sigma, \omega$, and $ \eta$ by the stage-two spin-
correlation
function $ I_4 $ for the production-decay sequence $ p \bar
p$
or $pp \rightarrow t \bar{t} X$ with $t\rightarrow W^{+} b$
and
$\bar t \rightarrow W^{-} \bar b$.  Sensitivities for
measurements of the polarized-partial-widths
$\Gamma_{L,T}^{\pm}$.  Sensitivities for the two tests for
\newline
``non-CKM-type" CP violation in $t\rightarrow W^{+}b$ decay.
Lastly, effective-mass scales $\Lambda$ in $TeV$ for ``new
physics" from additional Lorentz structures such as pure $V$
or $A$, or from a chiral combination of tensorial couplings
$g_+ \equiv f_M + f_E$ and  $g_- \equiv f_M - f_E$ [limits
are
given for purely real and for purely imaginary $g_i$'s].

\begin{center}
{\bf Figure Caption}
\end{center}

FIG. 1: The two pairs of spherical angles $\theta _1^t$,
$\phi
_1^t$ and  $\tilde \theta _a$%
, $\tilde \phi _a$ describe the two stages in the
sequential
decay $%
t\rightarrow W^{+}b$ followed by  $W^{+}\rightarrow j_{\bar
d}j_u$, or $%
W^{+}\rightarrow l^{+}\nu $.  The spherical angles
$\tilde \theta _a$, $%
\tilde \phi _a$ specify the $j_{\bar d}$ jet (or the $l^{+}$)
momentum in the $W^{+}
$rest frame when the boost is from the $t$ rest frame. For
the
hadronic $W^{+}$ decay mode, we use the notation that the
momentum of the charge $\frac{1}{3} e$ jet is denoted by
$j_{\bar d}$ and the momentum of the charge $\frac{2}{3} e$
jet by $j_{u}$.  In this figure, $\phi _1^t$ is
shown equal to
zero for simplicity of illustration.

\end{document}